\documentclass[12pt]{article}
\usepackage[cp1251]{inputenc}
\usepackage[english,russian]{babel}
\usepackage{indentfirst}
\usepackage{graphicx}
\usepackage{amsmath}
\usepackage{amssymb}
\input{epsf}

\newcommand{\eps}{\varepsilon}

\newcommand{\dd}{\partial}

\begin{document}

\begin{center}
\textbf{\Large Solitary Dust--Acoustic Waves in a Plasma with Two-Temperature Ions and Distributed Grain Size}
\\
\bigskip
\textbf{\large V. V. Prudskikh}
\\
\bigskip
 \textit{Research Institute of Physics, Southern Federal University, Rostov-on-Don, Russia}

\medskip

\end{center}

\noindent The propagation of weakly nonlinear dust--acoustic waves in a dusty plasma containing two ion species with different temperatures is explored. The nonlinear equations describing both the quadratic and cubic plasma nonlinearities are derived. It is shown that the properties of dust--acoustic waves depend substantially on the grain size distribution. In particular, for solitary dust--acoustic waves with a positive potential to exist in a plasma with distributed grain size, it is necessary that the difference between the temperatures of two ion species be large that that in the case of unusized grains.

\section{INTRODUCTION}

Dust--acoustic waves are a phenomenon inherent in plasmas containing, besides electrons and ions, heavy (about $10^{-14}$~g) dust grains, whose charges are determined by the balance between the ion and electron currents onto on the grain surface. The processes occurring in dusty plasmas have been actively studied over the past two decades. The presence of dust in plasma not only modifies the conventional plasma modes, but also gives rise to new types of low-frequency electrostatic plasma oscillations, such as dust--acoustic waves \cite{Rao1}, dust Coulomb waves \cite{Rao2}, and dust--lattice waves \cite{Mel1}. A considerable number of studies have been devoted to investigating linear and nonlinear properties of these waves. Weakly nonlinear small-amplitude waves are usually analyzed by expanding the equations in power series in a small parameter and using the method of coordinate stretching in accordance with the form of the dispersion relation \cite{Wash1}, while large-amplitude waves are studied using the quasi-potential method \cite{Sagd1} or gasdynamic approach \cite{Mac1}.

Dust--acoustic waves propagate with a very low phase velocity (from several tens centimeters per second up to a few meters per second). Therefore, it can be assumed that the electrons and ions, whose thermal velocities are much higher than the wave phase velocity, are distributed in the wave field according to the Boltzmann law. A number of papers were devoted to studying the properties of nonlinear dust--acoustic waves in a plasma containing two ion species with different temperatures \cite{Tag1}--\cite{Lab1}. Such a situation can occur when two plasmas with different temperatures are mixed with one another. The time during which ions relax to a state characterized by one temperature is usually much longer that the period of dust--acoustic oscillations (see \cite{Xie1}). Therefore, such a two-temperature plasma can be described by Boltzmann distributions for each ion species. It is also assumed that the electron temperatures of two plasmas are equalized very rapidly by virtue of the condition  $\tau^{ee}/\tau^{ii}\approx (m_e/m_i)^{1/2}(T_i/T_e)^{3/2}\ll 1$,
where $\tau^{ee}$ and $\tau^{ii}$ are the relaxation times of electrons and ions to equilibrium states; $m_e$ and $m_i$ are the masses of an electron and ion, respectively; $T_e$ and $T_i$ are the electron and ion temperatures.

In \cite{Tag1}, the propagation of a nonlinear dust--acoustic wave in a dusty plasma with two-temperature ions and a constant grain charge was considered. It was shown that both compression and rarefaction solitary waves, as well as double layers (waves having one nonzero asymptotics at infinity and described by the hyperbolic tangent function), can propagate in such a medium. In \cite{Xie1}, it was noticed that it is necessary to take into account variations of the grain charge in the wave field, because the contribution of this effect to the nonlinear coefficient in the Korteweg--de Vries (KdV) equation, describing the propagation of weakly nonlinear dust--acoustic waves, is comparable with the hydrodynamic nonlinearity.

In previous studies on the propagation of dust--acoustic waves in a two-temperature plasma, it was assumed that that dust grains are equal in size. However, it is well known that both laboratory and astrophysical dusty plasmas contain dust grains of different size \cite{Wan1}--\cite{Hor1}. The distribution function of dust grains over their radii $a$ can be described by the power law
\begin{equation}
f(a)=Ga^{-\nu},
\end{equation}
where $G$~--- is the normalizing factor. In calculations, it is usually assumed that the value of the exponent $\nu$ is in the range 3.5 -- 4 \cite{Cram1}; however, one can find in the literature alternative numbers. Thus, in \cite{Tri1} , the value of $\nu$ was assumed to be in the range 0.9 -- 4.5, and, in \cite{Ver1} , it was reported that, for Saturn's G rings, this parameter can even reach a value of 6--7. The model implies that the grain radii lie within the interval between the minimum value $a_1$ and the maximum value $a_2$; it is assumed that, outside this interval, $f(a)=0$.

In the present paper, the propagation of dust--acoustic waves in a two-temperature plasma is studied with allowance for both adiabatic variations in the grain charge (caused by variations in the densities of the ambient electrons and ions in the wave field and, accordingly, variations in the fluxes of negative and positive charges onto the grain surface) and changes in distribution function (1). The size distribution function alters both the linear wave phase velocity (as compared to the case of a dust with unisized grains) and the nonlinear coefficient in the KdV equation. An important point is that this coefficient can change its sign. Previous studies have shown that it can take both negative and positive values, which corresponds to compression and rarefaction dust waves, respectively. The latter is possible when the ratio between the densities of cold and hot ions is moderate, whereas their temperatures differ significantly. The goal of the present study was to investigate the propagation of solitary dust--acoustic waves in a plasma with two-temperature ions under the assumption that dust grain sizes obey distribution (1), to find the conditions required for the existence of waves with a negative and positive potential, and to compare the results obtained with results of calculations performed for a dust with equal-size grains.

The paper is organized as follows. In Section 2, a set of basic equations is derived and dimensionless variables are introduced. In Section 3, the KdV equation and modified KdV equation (mKdV) with coefficients that take into account the distribution over dust grain sizes are obtained by expanding the basic equations in power series in a small parameter and using the method of coordinate stretching. In Section 4, the nonlinear coefficients of the equations obtained are analyzed. The main results are summarized in Section 5.

\section{BASIC EQUATIONS}

Let us consider a dusty plasma consisting of electrons, ions, and heavy dust grains of different size and mass. We assume that the dust contains $N$ different species with the densities $n_{j0}$, charges $Z_{j0}$, and masses $m_j$. Then, the condition of plasma quasi--neutrality can be written as
\begin{equation}
n_{ih0}+n_{il0}=n_{e0}+\sum^N_{j=1}Z_{j0}n_{j0},
\end{equation}
where $n_{ih0},n_{il0}$, and $n_{e0}$ are the unperturbed densities of high- and low-temperature ions and elect-\\rons, respectively. We assume that the wave phase is much larger than the dust thermal velocity and ignore the influence of the thermal pressure on the dynamic of the dust component. In this case, the continuity equation and equation of motion for each dust species have the form
\begin{equation}
\frac{\dd n_j}{\dd t}+\frac{\dd}{\dd x}\left(n_ju_j\right)=0,
\end{equation}
\begin{equation}
\frac{\dd u_j}{\dd t}+u_j\frac{\dd u_j}{\dd x}=\frac{Z_je}{m_j}\frac{\dd\phi}{\dd x},
\end{equation}
where $n_j$ and $u_j$~--- are the density and velocity of the $j$th dust component and $Z_j$~--- is the charge of its grains. This set of equations should be supplemented with Poisson's equation
\begin{equation}
\frac{\dd^2 \phi}{\dd x^2}=4\pi e(\sum^N_{j=1}Z_{j}n_{j}+n_{e}-n_{ih}-n_{il}),
\end{equation}
in which the densities of electrons and both ion components obey Boltzmann distributions,
\begin{equation}
n_e=n_{e0}\exp\left(\frac{e\phi}{T_e}\right),
\end{equation}
\begin{equation}
n_{ih}=n_{ih0}\exp\left(-\frac{e\phi}{T_{ih}}\right),
\end{equation}
\begin{equation}
n_{il}=n_{il0}\exp\left(-\frac{e\phi}{T_{il}}\right).
\end{equation}
Here, $T_e, T_{ih}$, and $T_{il}$ are the temperatures of electrons and high- and low-temperature ions, respective-\\ly.

Variations in the charge of dust grains of the $j$th species are described by the balance equation of currents onto the grain surface,
\begin{equation}
e\frac{dZ_j}{dt}=I_e+I_{ih}+I_{il},
\end{equation}
where
\begin{displaymath}
I_e=-e\pi a_j^2(8T_e/m_e)^{1/2}n_e\exp\left(\frac{e\Phi_j}{T_e}\right),
\end{displaymath}
\begin{displaymath}
I_{ih}=e\pi a_j^2(8T_{ih}/m_i)^{1/2}n_{ih}\left(1-\frac{e\Phi_j}{T_{ih}}\right),
\end{displaymath}
\begin{displaymath}
I_{il}=e\pi a_j^2(8T_{il}/m_i)^{1/2}n_{il}\left(1-\frac{e\Phi_j}{T_{il}}\right),
\end{displaymath}
$\Phi_j=-Z_je/a_j$ is the surface potential of grains of the $j$th dust component and $a_j$ is their radius. The charging frequency of dust is determined by the expression $\nu_{ch}=a_j\omega_{pi}/(\sqrt{2\pi}\lambda_{Di})$, where $\omega_{pi}=(4\pi e^2m_{i0}/m_i)^{1/2}$ is the ion plasma frequency and $\lambda_{Di}=(T_i/4\pi e^2n_{i0})^{1/2}$ is the ion Debye length. The time during which a wave passes through a given point of the medium is $\tau_d=\lambda/v_{ph}\approx(\lambda/\lambda_{Di})\omega_{pd}^{-1}$,
where $\omega_{pd}=(4\pi Z^2e^2n_{d0}/m_d)^{1/2}$ and $\lambda$ is the wavelength. We assume that the charge of a dust grain in the wave field varies quasi-statically, so the current balance equation has the form
\begin{equation}
I_e+I_{ih}+I_{il}\approx 0.
\end{equation}
Obviously, for this to occur, it is necessary that $\nu_{ch}\tau_d\gg 1$.  Assuming that  $a_j\sim 10^{-4}$~cm, $T_i\sim 1$~eV, $Z\sim 10^3$, $m_d\sim
10^{-14}$~г, and $Zn_{d0}\sim n_{i0}$, we obtain the following condition under which the dust grain charge in a long-wavelength dust--acoustic wave $\lambda_{Di}\ll\lambda$  can be described adiabatically \cite{Ma1}: $n_{i0}\gg 10^6(\lambda_{Di}/\lambda)$~cm$^{-3}$. For this condition to be satisfied, the plasma should be sufficiently dense. Otherwise, it is necessary to take into account the nonadiabatic character of grain changing in the wave field, due to which the initial perturbation transforms into a dust--acoustic shock wave \cite{Gup1}. It follows from the above estimate that the model in which the grain charge is assumed to vary adiabatically mainly applies to laboratory plasmas, while the use of this model to describe dust--acoustic waves in astrophysical plasmas, e.g., in Saturn's F rings (\cite{Lab1}), is incorrect.

In equation of dust motion (4), we ignored the frictional force. In dusty plasmas, the momentum loss of dust grains due to collisions with ions is usually negligibly small and can be ignored. Actually, the applicability condition of this approach is $\varkappa=m_in_{i0}\nu_{id}/(m_dn_{d0}\omega_{pd})\ll 1$, where $\nu_{id} =\\4\pi\Lambda Z^2e^4n_{d0}/m_i^{1/2}T_i^{3/2}$ is the collision frequency of ions with dust grains and  $\Lambda$ is the Coulomb logarithm. For the above parameter values and $n_{d0}\sim
10^6$~см$^{-3}$, we have $\varkappa\sim 10^{-4}$. In laboratory plasmas, the density of neutral particles is, as a rule, several orders of magnitude higher than the plasma density. In our problem, the condition under which the friction of dust by the neutral component can be ignored has the form $\nu_{dn}/\omega_{pd}=8\sqrt{2\pi}a_j^2m_nn_nv_{Tn}/3m_d\omega_{pd}\ll
1$ \cite{Shuk1}, where $n_n$ is the density of neutral particles, $m_n$ is the mass of a neutral atom, and $v_{Tn}$ is their thermal velocity. For $n_{d0}$, we find that the density of neutral particles should be $n_n\ll 10^{15}$~см$^{-3}$, which is usually satisfied with a large margin.

Let us introduce the total dust density $N=\sum^N_{j=1}n_{j0}$ and the average charge  $\overline{Z_0}=\\=\sum^N_{j=1}n_{j0}Z_{j0}/N$, average radius $\overline{a}=\sum^N_{j=1}a_jn_{j0}/N$ , and average mass $\overline{m}=\sum^N_{j=1}m_jn_{j0}/N$ of a dust grain. We also introduce the quantities
\begin{displaymath}
T_{eff}=\frac{\overline{Z_0}NT_eT_{ih}T_{il}}{n_{e0}T_{ih}T_{il}+n_{ih0}T_{e}T_{il}+n_{il0}T_{e}T_{ih}},
\end{displaymath}
$\lambda_{Dd}=(T_{eff}/4\pi \overline{Z_0}e^2N)^{1/2}$,
$\omega_{pd}=(4\pi \overline{Z_0}^2e^2N/\overline{m})^{1/2}$, and
$c_D=(\overline{Z_0}T_{eff}/\overline{m})^{1/2}$, and normalize the dust density $n_j$ to $N$; the dust grain charge $Z_j$ to $\overline{Z_0}$; the grain mass $m_j$ to $\overline{m}$; the electron density $n_{e0}$ and ion densities $n_{il0}, n_{ih0}$ to $\overline{Z_0}N=\sum^N_{j=1}n_{j0}Z_{j0}$; the time $t$ to $\omega_{pd}^{-1}$, the spatial coordinate $x$ to $\lambda_{Dd}$; the velocity $u_j$ to $c_D$; and the electrostatic potential $\phi$ and surface potential $\Phi_j$ to $T_{eff}/e$. Then, the initial set of equations takes the form
\begin{equation}
\frac{\dd n_j}{\dd t}+\frac{\dd}{\dd x}\left(n_ju_j\right)=0,
\end{equation}
\begin{equation}
\frac{\dd u_j}{\dd t}+u_j\frac{\dd u_j}{\dd x}=\frac{Z_j}{m_j}\frac{\dd\phi}{\dd x},
\end{equation}
\begin{equation}
\frac{\dd^2 \phi}{\dd x^2}=\sum^N_{j=1}Z_{j}n_{j}+\frac{1}{\delta_1+\delta_2-1}(\exp(\beta_1s\phi)-\delta_1\exp(-s\phi)-\delta_2\exp(-\beta s\phi)),
\end{equation}
\begin{equation}
\exp[\beta_1s(\phi+\Phi_j)]=\alpha_1\delta_1\exp(-s\phi)(1-s\Phi_j)+\alpha_2\delta_2\exp(-\beta s\phi)(1-\beta s\Phi_j),
\end{equation}
where the following notation is introduced: $\delta_1=n_{il0}/n_{e0}$, $\delta_2=n_{ih0}/n_{e0}$,
$\beta_1=T_{il}/T_e$, $\beta_2=T_{ih}/T_e$,
$\beta=T_{il}/T_{ih}=\beta_1/\beta_2$,
$\alpha_1=(\beta_1/\mu)^{1/2}$, $\alpha_2=(\beta_2/\mu)^{1/2}$,
$\mu=m_i/m_e\approx 1836$, and $s=T_{eff}/T_{il}=(\delta_1+\delta_2-1)/(\beta_1+\delta_1+\beta\delta_2)$.

In deriving Eq. (13), we used the condition $\sum^N_{j=1}n_{j0}Z_{j0}=1$, written in dimensionless variables.

\section{WEAKLY NONLINEAR APPROXIMATION FOR DUST--ACOUSTIC WAVES}

We will search for the solution to the set of Eqs. (11)--(14) in the form
\begin{equation}
\mathbf{Y}=\mathbf{Y}_0+\eps
\mathbf{Y}_1+\eps^2\mathbf{Y}_2+\eps^3\mathbf{Y}_3+\ldots,
\end{equation}
where $\mathbf{Y} = (\phi, n_j, u_j, \Phi_j)$ and $\mathbf{Y}_0 =
(0,n_{j0},0,\Phi_{j0})$, by using the conventional method of coordinate stretching:
\begin{equation}
\xi=\eps^{1/2}(x-V_0t),\hspace{1 cm}\tau=\eps^{3/2}t.
\end{equation}

In the zeroth order in $\eps$, we obtain the equation determining the surface potential of dust grains in an unperturbed plasma,
\begin{equation}
\exp(\beta_1s\Phi_{j0})=\alpha_1\delta_1(1-s\Phi_{j0})+\alpha_2\delta_2(1-\beta s\Phi_{j0}).
\end{equation}
Since, in this equation, the surface potential $\Phi_{j0}$ of different species of dust grains does not depend on $j$, the subscript $j$ in $\Phi_{j0}$ will be further omitted.

In the first order in $\eps$, the set of equations has the form
\begin{equation}
-V_0\frac{\dd n_{j1}}{\dd\xi}+n_{j0}\frac{\dd u_{j1}}{\dd\xi}=0,
\end{equation}
\begin{equation}
V_0\frac{\dd u_{j1}}{\dd\xi}+\frac{Z_{j0}}{m_j}\frac{\dd \phi_1}{\dd\xi}=0,
\end{equation}
\begin{equation}
\sum^N_{j=1}Z_{j1}n_{j0}+\sum^N_{j=1}Z_{j0}n_{j1}+\phi_1=0.
\end{equation}

Integrating Eqs. (18)--(20) with the boundary conditions $\phi_1,
n_{j1}, u_{j1}, Z_{j1}\rightarrow 0$ at $\xi\rightarrow\pm\infty$,
we obtain the following relationships for the lowest order quantities:
\begin{equation}
n_{j1}=-\frac{Z_{j0}n_{j0}}{m_jV_0^2}\phi_1,\hspace{0.5 cm} u_{j1}=-\frac{Z_{j0}}{m_jV_0}\phi_1,\hspace{0.5 cm}Z_{j1}=\gamma_1\phi_1,
\end{equation}
where the factor $\gamma_1$ is defined in the Appendix. For the linear wave velocity, we have
\begin{equation}
V_0^2=\cfrac{\sum^N_{j=1}\cfrac{Z_{j0}^2n_{j0}}{m_j}}{1+\gamma_1}.
\end{equation}

In the second order in $\eps$, the set of equations takes the form
\begin{equation}
-V_0\frac{\dd n_{j2}}{\dd\xi}+\frac{\dd n_{j1}}{\dd\tau}+n_{j0}\frac{\dd u_{j2}}{\dd\xi}+\frac{\dd}{\dd\xi}(n_{j1}u_{j1})=0,
\end{equation}
\begin{equation}
-V_0\frac{\dd u_{j2}}{\dd\xi}+\frac{\dd u_{j1}}{\dd\tau}+u_{j1}\frac{\dd u_{j1}}{\dd\xi}- \frac{Z_{j0}}{m_j}\frac{\dd \phi_2}{\dd\xi}-\frac{Z_{j1}}{m_j}\frac{\dd \phi_1}{\dd\xi}=0,
\end{equation}
\begin{equation}
\frac{\dd^2 \phi_1}{\dd \xi^2}=\sum^N_{j=1}Z_{j2}n_{j0}+\sum^N_{j=1}Z_{j1}n_{j1}+\sum^N_{j=1}Z_{j0}n_{j2}+ \frac{s^2(\beta_1^2-\delta_1-\beta^2\delta_2)}{2(\delta_1+\delta_2-1)}\phi_1^2+\phi_2,
\end{equation}
\begin{equation}
Z_{j2}=\gamma_1\phi_2+\gamma_2\phi_1^2,
\end{equation}
where the factor $\gamma_2$ is defined in the Appendix.

Expressing $\dd n_{j2}/\dd\xi$ from Eqs. (23) and (24) and using Eq. (21), we obtain the KdV equation for a weakly nonlinear dust--acoustic wave,
\begin{equation}
\frac{\dd\phi_1}{\dd\tau}+B\phi_1\frac{\dd\phi_1}{\dd\xi}+A\frac{\dd^3\phi_1}{\dd\xi^3}=0.
\end{equation}
The coefficients $A$ and $B$ in this equation are
\begin{displaymath}
A=\frac{V_0}{2(1+\gamma_1)},\hspace{0.5 cm} B/A=\frac{s^2(\delta_1+\beta^2\delta_2-\beta_1^2)}{\delta_1+\delta_2-1}-\frac{3}{V_0^4}\sum_{j=1}^N\frac{Z_{j0}^3n_{j0}}{m_j^2}+\frac{3\gamma_1}{V_0^2}\sum_{j=1}^N\frac{Z_{j0}n_{j0}}{m_j}-2\gamma_2.
\end{displaymath}

To analyze the case in which the ratio between the coefficients in Eq. (27) is zero,  $B/A=0$, we use the method of coordinate stretching,
\begin{equation}
\xi=\eps(x-V_0t),\hspace{1 cm}\tau=\eps^{3}t.
\end{equation}
Then, the time derivative drops out from Eqs. (23) and (24) and we obtain the following expressions for the second-order quantities:
\begin{equation}
n_{j2}=\frac{n_{j0}}{m_jV_0^2}\left[-Z_{j0}\phi_2+\frac{1}{2}\left(\frac{3Z_{j0}^2}{m_jV_0^2}-\gamma_1\right)\phi_1^2\right],\nonumber
\end{equation}
\begin{equation}
u_{j2}=\frac{1}{m_jV_0}\left[-Z_{j0}\phi_2+\frac{1}{2}\left(\frac{Z_{j0}^2}{m_jV_0^2}-\gamma_1\right)\phi_1^2\right].
\end{equation}
In the third order in $\eps$, the set of Eqs. (11)--(14) takes the form
\begin{equation}
-V_0\frac{\dd n_{j3}}{\dd\xi}+\frac{\dd n_{j1}}{\dd\tau}+n_{j0}\frac{\dd u_{j3}}{\dd\xi}+\frac{\dd}{\dd\xi}(n_{j1}u_{j2}+n_{j2}u_{j1})=0,
\end{equation}
\begin{equation}
-V_0\frac{\dd u_{j3}}{\dd\xi}+\frac{\dd u_{j1}}{\dd\tau}+\frac{\dd }{\dd\xi}(u_{j1}u_{j2})- \frac{Z_{j0}}{m_j}\frac{\dd \phi_3}{\dd\xi}-\frac{Z_{j1}}{m_j}\frac{\dd \phi_2}{\dd\xi}-\frac{Z_{j2}}{m_j}\frac{\dd \phi_1}{\dd\xi}=0,
\end{equation}
\begin{gather}
\frac{\dd^2 \phi_1}{\dd \xi^2}=\sum^N_{j=1}Z_{j3}n_{j0}+\sum^N_{j=1}Z_{j2}n_{j1}+\sum^N_{j=1}Z_{j1}n_{j2}+\sum^N_{j=1}Z_{j0}n_{j3}+\frac{s^3(\beta_1^3+\delta_1+\beta^3\delta_2)}{6(\delta_1+\delta_2-1)}\phi_1^2+\nonumber\\
+\frac{s^2(\beta_1^2-\delta_1-\beta^2\delta_2)}{\delta_1+\delta_2-1}\phi_1\phi_2+\phi_3,
\end{gather}
\begin{equation}
Z_{j3}=\gamma_1\phi_3+2\gamma_2\phi_1\phi_2+\gamma_3\phi_1^3,
\end{equation}
where the factor $\gamma_3$ is defined in the Appendix.

Excluding $\dd u_{j3}/\dd\xi$ from Eqs. (30) and (31), differentiating expression (32) over $\xi$, and using Eqs. (21) and (29) and the condition $B/A=0$,
we arrive at the mKdV equation,
\begin{equation}
\frac{\dd\phi_1}{\dd\tau}+C\frac{\dd\phi_1^3}{\dd\xi}+A\frac{\dd^3\phi_1}{\dd\xi^3}=0,
\end{equation}
where
\begin{gather}
C/A=\frac{5}{2V_0^6}\sum_{j=1}^N\frac{Z_{j0}^4n_{j0}}{m_j^3}-\frac{3\gamma_1}{V_0^4}\sum_{j=1}^N\frac{Z_{j0}^2n_{j0}}{m_j^2}+\frac{\gamma_1^2}{2V_0^2}\sum_{j=1}^N\frac{n_{j0}}{m_j}+\nonumber\\
+\frac{4\gamma_2}{3V_0^2}\sum_{j=1}^N\frac{Z_{j0}n_{j0}}{m_j}-\gamma_3-\frac{s^3}{6}\frac{\beta_1^3+\delta_1+\beta^3\delta_2}{\delta_1+\delta_2-1}.\nonumber
\end{gather}

Under the condition $\phi_1(\eta=\xi-u\tau\rightarrow\pm\infty)=0$ , Eqs. (27) and (34) have the well-known solutions in the form of solitary waves. Their properties, such as the amplitude, the propagation velocity, and the sign of the electric potential, are determined by the values and signs of the coefficients entering into the nonlinear equations. Below, we will analyze the dependence of these coefficients on the plasma parameters.

\section{ANALYSIS OF NONLINEAR COEFFICIENTS}

Taking into account distribution function (1), we replace the sum over dust species with the integral,
\begin{equation}
\sum_{j=1}^Nn_{j0}=G\int_{a_{1}}^{a_{2}}a^{-\nu}da=N.
\end{equation}
Then, averaging all the quantities under the summation symbol in the expressions for  $V_0^2$, $B/A$ and $C/A$, we obtain
\begin{equation}
V_0^2=\frac{v_1}{1+\gamma_1},
\end{equation}
\begin{equation}
B/A=\frac{s^2(\delta_1+\beta^2\delta_2-\beta_1^2)}{\delta_1+\delta_2-1}-\frac{3b_1}{V_0^4}+\frac{3\gamma_1b_2}{V_0^2}-2\gamma_2,
\end{equation}
\begin{equation}
C/A=\frac{5c_1}{2V_0^6}-\frac{3\gamma_1c_2}{V_0^4}+\frac{\gamma_1^2c_3}{2V_0^2}
+\frac{4\gamma_2c_4}{3V_0^2}-\gamma_3-\frac{s^3}{6}\frac{\beta_1^3+\delta_1+\beta^3\delta_2}{\delta_1+\delta_2-1},
\end{equation}
where
\begin{gather}
v_1=\frac{(2-\nu)^2}{\nu(4-\nu)}\frac{(a_2^{4-\nu}-a_1^{4-\nu})(a_2^{-\nu}-a_1^{-\nu})}{(a_2^{2-\nu}-a_1^{2-\nu})^2},\nonumber\\
b_1=\frac{(2-\nu)^3}{(4-\nu)^2(2+\nu)}\frac{(a_2^{4-\nu}-a_1^{4-\nu})^2(a_1^{-2-\nu}-a_2^{-2-\nu})}{(a_2^{2-\nu}-a_1^{2-\nu})^3},\nonumber\\
b_2=\frac{(2-\nu)(1-\nu)}{(4-\nu)(1+\nu)}\frac{(a_2^{4-\nu}-a_1^{4-\nu})^2(a_1^{-1-\nu}-a_2^{-1-\nu})}{(a_2^{2-\nu}-a_1^{2-\nu})(a_2^{1-\nu}-a_1^{1-\nu})},\\
c_1=\frac{(2-\nu)^4}{(4-\nu)^3(4+\nu)}\frac{(a_2^{4-\nu}-a_1^{4-\nu})^3(a_1^{-4-\nu}-a_2^{-4-\nu})}{(a_2^{2-\nu}-a_1^{2-\nu})^4},\nonumber\\
c_2=\frac{(2-\nu)^2(1-\nu)}{(4-\nu)^2(3+\nu)}\frac{(a_2^{4-\nu}-a_1^{4-\nu})^2(a_1^{-3-\nu}-a_2^{-3-\nu})}{(a_2^{2-\nu}-a_1^{2-\nu})^2(a_2^{1-\nu}-a_1^{1-\nu})},\nonumber\\
c_3=\frac{(1-\nu)^2}{(4-\nu)(2+\nu)}\frac{(a_2^{4-\nu}-a_1^{4-\nu})(a_1^{-2-\nu}-a_2^{-2-\nu})}{(a_2^{1-\nu}-a_1^{1-\nu})^2},\nonumber\\
c_4=b_2\nonumber.
\end{gather}
In deriving the above expressions, we took into account that the density $n_{j0}$, charge $Z_{j0}$, and mass $m_j$ are dimensionless quantities; therefore, in taking integrals, we passed to dimensional variables in accordance with $n\rightarrow n/N$, $Z(a)\rightarrow Z(a)/\overline{Z_0}$, and $m(a)\rightarrow m(a)/\overline{m}$. We also used the dependences  $Z(a)=(\Phi_0/e)a$, and $m(a)=(4\pi\rho/3) a^3$ to find expressions for the average values of $\overline{Z_0}$ and $\overline{m}$:
\begin{equation}
\overline{Z_0}=\frac{\Phi_0}{e}\frac{1-\nu}{2-\nu}\frac{a_2^{2-\nu}-a_1^{2-\nu}}{a_2^{1-\nu}-a_1^{1-\nu}},\hspace{0.5 cm}\overline{m}=\frac{4\pi\rho}{3}\frac{1-\nu}{4-\nu}\frac{a_2^{4-\nu}-a_1^{4-\nu}}{a_2^{1-\nu}-a_1^{1-\nu}},
\end{equation}
using which expressions (39) can easily be derived.

Formulas (39) allow a limiting transition to the case of a dust with unisized grains at  $a_2\rightarrow a_1$, or $\nu\rightarrow\infty$. In the former case, the uncertainty is eliminated by setting  $a_2=a_1+\Delta a$ with $\Delta a\ll a$ and keeping the summands of the first order of smallness. In the latter case, the contribution from the distribution function is significant only in the vicinity of $a_1$ , irrespective of the width of the interval $[a_1-a_2]$. In both cases, coefficients (39) are reduced to unity and nonlinear coefficients (37) and (38) transform into the expressions obtained in \cite{Xie1}.

The results of numerical analysis of coefficients (37) and (38) are presented in Figs. 1--6. Figure 1 shows the dependence of the coefficient $B/A$ on $\delta_2$ for two values of $\delta_1$ for the case of a dust of unusized grains. This coefficient can vanish when the density of cold ions is smaller than the density of hot ions and the temperature ratio  $T_{ih}/T_{il}$ is high enough. As the absolute value of the parameter $\delta_1$ (which characterizes the density of cold ions) increases, the zeros of the function $B/A(\delta_1)$ shift to the right. Figure 2 illustrates the dependence of the coefficient $B/A$ on $\delta_2$ for different ratios $a_2/a_1$ and a fixed value of the exponent $\nu$. It is seen that, as the width of the interval of dust grain radii increases, the critical point $B/A=0$ disappears. Figure 3 presents a series of plots $B/A(\delta_1)$ $a_2/a_1$ and different values of the exponents $\nu$. As was mentioned above, the case of large $\nu$ values corresponds to a dust with grains nearly equal in size. For the given parameters, the coefficient $B/A$ at
$\nu\rightarrow\infty$ can be either negative or positive. However, as the exponent $\nu$ decreases, the influence of the distributed grain size increases, the critical point disappears, and the coefficient becomes negative defined. Figure 4 shows the dependence $B/A(\delta_2)$ for different ratios between the temperatures of cold and hot ions. It is seen that the coefficient is very sensitive to the ion temperature ratio. For the given (wide) interval of dust grain radii, solitary rarefaction waves can exist only when the temperature ratio is larger than 35. At the same time, in a dusty plasma with equal-size grains, rarefaction waves can exist when the temperature ratio is 20, the other parameters being the same (see Fig. 1).

The curves in the plane $(\delta_1 - \delta_2)$ in Fig. 5 correspond to the points at which the coefficient $B/A$ vanishes. The lower and upper curves correspond to the left and right zeros of the function $B/A$. For $\delta_1$ values in the range $[0-0.85]$, there are two points on curve 1 at which the nonlinear coefficient vanishes. The existence of two critical points of this coefficient in a dusty plasma with two-temperature ions was revealed in \cite{Tag1}. At $\delta_1=0.85$, the critical points merge, and, at $\delta_1>0.85$, the coefficient is negative defined. The region in which $B/A>0$ (rarefaction dust wave) lies inside the sector bounded by the curve. The larger the spread in dust grain sizes, the narrower the domain of existence of rarefaction waves (see curve 2). At $\nu<2.5$ (the other parameters being the same), the coefficient $B/A$ is negative defined and nonlinear dust--acoustic waves are compression waves. The dependence $C/A(\delta_1)$ under the condition $B/A=0$
has the shape of a loop (see Fig. 6). The two points at which the coefficient $B/A$ vanishes correspond to two different curves $C/A$. Curves \textit{1} and \textit{2} refer to the left and right zeros of $B/A$, respectively. At $\delta_1\approx 0.85$ , the zeros of the coefficient $B/A$ merge; accordingly, the values of $C/A$ also merge at that point. It should be noted that, on curve \textit{2}, the coefficient $C/A$ can take negative values. This means that double layer structures can form in a dusty plasma. We also note that, at $C/A=0$ (curve \textit{2}) , the second-order nonlinearities are mutually cancelled. In this case, the nonlinear properties of a solitary wave are determined by the third-order terms and can be analyzed in a similar way \cite{Tag1}.

\section{CONCLUSIONS}

As the range $[a_1,a_2]$ of dust grain sizes extends and/or the exponent $\nu$ of distribution functions (1) decreases, the positive values of the nonlinear coefficient $B/A$ n the KdV equation decrease and its negative values increase in magnitude (see Figs. 2, 3). In this case, the amplitude of a rarefaction solitary wave increases and that of a compression solitary waves decreases as compared to those in a dusty plasma with unisized grains. Even if the ratio between the temperatures of cold and hot ions, $T_{il}/T_{ih}$, , is quite sufficient for the development of solitary waves with a positive potential in a dusty plasma with unisized grains, it can be insufficient for their development in the case where the grain sizes obey distribution (1) (cf. curve 1 in Fig. 1 with curve 3 in Fig. 4). Thus, in order for rarefaction solitary dust--acoustic waves to exist in a plasma with two-temperature ions and distributed dust grain size, it is necessary that the temperature ratio between two ion components be larger than that in a dusty plasma with unisized grains.

\section{ACKNOWLEDGMENTS}
The author is grateful to Yu.A. Shchekinov for helpful discussions. This work was supported by the Russian Federal Agency on Education (project no. RNP 2.1.1.3483) and the Russian Foundation for Basic Research (project nos. 05-02-17070 and 06-02-16819).

\section*{APPENDIX}

Let us define constants relating perturbations of the dust grain charge to the wave potential $\phi$. We expand the equation
\begin{equation}
\exp[\beta_1s(\phi+\Phi)]=\alpha_1\delta_1\exp(-s\phi)(1-s\Phi)+\alpha_2\delta_2\exp(-\beta s\phi)(1-\beta s\Phi)\nonumber
\end{equation}
in power series in $\phi$ and $\Phi$. In the first order in order $\eps$, we have
\begin{gather}
\phi_1[\beta_1\exp[\beta_1s\Phi_0)+\alpha_1\delta_1(1-s\Phi_0)+\alpha_2\delta_2\beta(1-\beta s\Phi_0)]+\nonumber\\
+\Phi_1[\beta_1\exp[\beta_1s\Phi_0)+\alpha_1\delta_1+\alpha_2\delta_2\beta]=0.\nonumber
\end{gather}
Introducing the quantities
\begin{gather}
\Gamma_1=\beta_1\exp[\beta_1s\Phi_0)+\alpha_1\delta_1(1-s\Phi_0)+\alpha_2\delta_2\beta(1-\beta s\Phi_0)\nonumber\\
\Gamma_2=\beta_1\exp[\beta_1s\Phi_0)+\alpha_1\delta_1+\alpha_2\delta_2\beta,\nonumber
\end{gather}
we obtain
\begin{equation}
\Phi_1=-\frac{\Gamma_1}{\Gamma_2}\phi_1.\nonumber
\end{equation}
Using the relationship $Z_1=\Phi_1/\Phi_0$ for dimensionless quantities, we find
\begin{equation}
Z_1=\gamma_1\phi_1,\text{ where }\gamma_1=-\frac{\Gamma_1}{\Gamma_2\Phi_0}.\nonumber
\end{equation}
The factor $\gamma_1$  is the same for any $j$th dust species; therefore, the subscript $j$ in  $\Phi_{jk}$ and $Z_{jk}$ ($k=1,2,3$) can be omitted. 

In the second order in $\eps$, we have
\begin{gather}
\Gamma_1\phi_2+\Gamma_2\Phi_2=-\frac{1}{2}\beta_1^2s\exp(\beta_1s\Phi_0)(\phi_1+\Phi_1)^2+\alpha_1\delta_1s(\phi_1\Phi_1+(1-s\Phi_0)\phi_1^2/2)+\nonumber\\
+\alpha_2\delta_2\beta^2s(\phi_1\Phi_1+(1-\beta s\Phi_0)\phi_1^2/2)\nonumber.
\end{gather}
Using the relationships $\Phi_1=\gamma_1\Phi_0\phi_1$ and $Z_2=\Phi_2/\Phi_0$, we obtain
\begin{gather}
Z_2=\gamma_1\phi_2+\gamma_2\phi_1^2,
\text{ where }\gamma_2=\frac{\Gamma_3}{\Gamma_2\Phi_0},\nonumber\\
\Gamma_3=-\frac{1}{2}\beta_1^2s\exp(\beta_1s\Phi_0)(1+\gamma_1\Phi_0)^2+\alpha_1\delta_1s(\gamma_1\Phi_0+(1-s\Phi_0)/2)+\nonumber\\
+\alpha_2\delta_2\beta^2s(\gamma_1\Phi_0+(1-\beta s\Phi_0)/2)\nonumber.
\end{gather}

Finally, in the third order in $\eps$, we obtain
\begin{gather}
\Gamma_1\phi_3+\Gamma_2\Phi_3=2\Gamma_3\phi_1\phi_2+\Gamma_4\phi_1^3,\nonumber\\
Z_3=\gamma_1\phi_3+2\gamma_2\phi_1\phi_2+\gamma_3\phi_1^3\nonumber
\end{gather}
where
\begin{gather}
\gamma_3=\frac{\Gamma_4}{\Gamma_2\Phi_0}, \Gamma_4=-\beta_1^2s(1+\gamma_1\Phi_0)\exp(\beta_1s\Phi_0) (\gamma_2\Phi_0+\beta_1s(1+\gamma_1\Phi_0)^2/6)-\nonumber\\ -\alpha_1\delta_1s^2(\gamma_1\Phi_0/2-\gamma_2\Phi_0/s+(1-s\Phi_0)/6)-\nonumber\\
-\alpha_2\delta_2\beta^3s^2(\gamma_1\Phi_0/2-\gamma_2\Phi_0/\beta s+(1-\beta s\Phi_0)/6)\nonumber.
\end{gather}

\newpage
\section*{FIGURE CAPTIONS}
Fig. 1. Nonlinear coefficient $B/A$ as a function of $\delta_2$ for the case of a dust with equal-size grains at different temperatures of the cold ion component: (a) $\beta_1$= 1) 0.03, 2) 0.05,  3) 0.07 ($\delta_1=1$, $\beta_2=1$); (b) $\beta_1$= 1) 0.02, 2) 0.03, 3) 0.05 ($\delta_1=2$, $\beta_2=1$). 
Key: 1. (a); 2. (b)
\vspace{1 cm}

Fig. 2. Nonlinear coefficient $B/A$ as a function $\delta_2$ for different values of the ratio $a_2/a_1$: 1) 1, 2) 1.5, 3) 2, and 4) 2.5 ($\delta_1=1$, $\beta_1=0.04$, $\beta_2=1$, $\nu=3.5$).
 \vspace{1 cm}

Fig. 3. Nonlinear coefficient $B/A$ as a function of $\delta_2$ for different values of $\nu$: 1) 7, 2) 6, 3) 5 ($\delta_1=1$,
$\beta_1=0.04$, $\beta_2=1$, $a_2/a_1=10$). \vspace{1 cm}

Fig. 4. Nonlinear coefficient $B/A$ as a function of $\delta_2$ for different values of $\beta_1$: 1) 0.02, 2) 0.025, 3) 0.03 ($\delta_1=1$, $\beta_2=1$, $a_2/a_1=10$, $\nu=3.5$). \vspace{1 cm}

Fig. 5. Curves defined by the condition $B/A=0$  in the plane $[\delta_1 - \delta_2]$ for $\nu$= 1) 3.5, 2) 3 ($\beta_1=0.03$, $\beta_2=1$, $a_2/a_1=10$).
\vspace{1 cm}

Fig. 6. Coefficient $C/A$ as a function of $\delta_1$ for $B/A=0$.
Curves 1 and 2 correspond to left and right zeros of the coefficient $B/A$,
respectively ($\beta_1=0.03$,
$\beta_2=1$, $a_2/a_1=10$, $\nu=3.5$).

\newpage

\begin{figure}\center
\includegraphics[height=6cm]{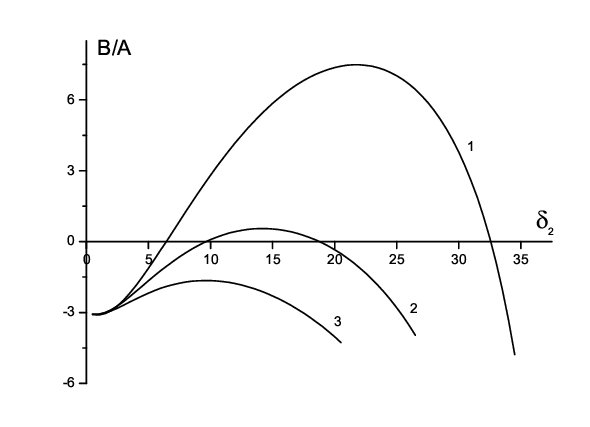}
\includegraphics[height=6cm]{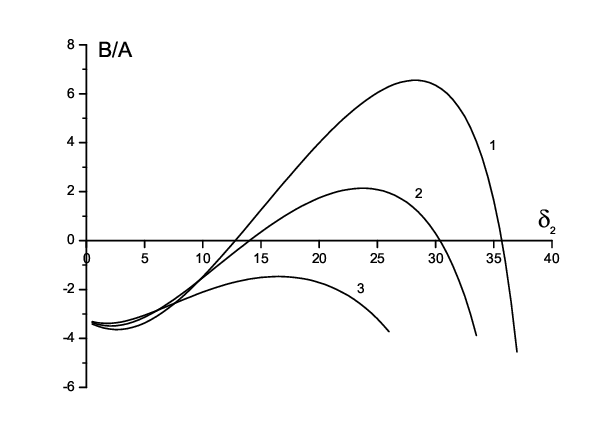}
\\Fig.1
\end{figure}

\begin{figure}\center
\includegraphics[height=6cm]{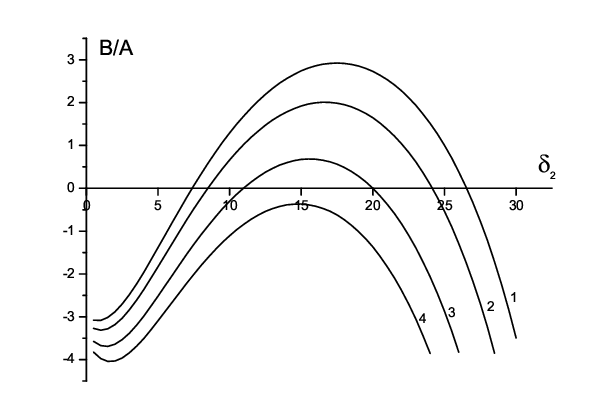}
\\Fig.2
\end{figure}

\begin{figure}\center
\includegraphics[0,200][300,300]{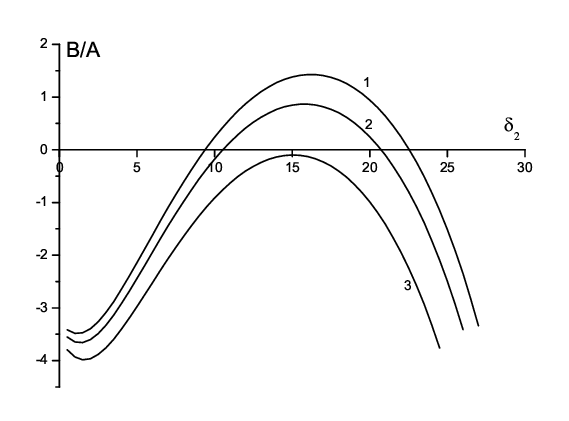}
\\Fig.3
\end{figure}

\begin{figure}\center
\includegraphics[height=6cm]{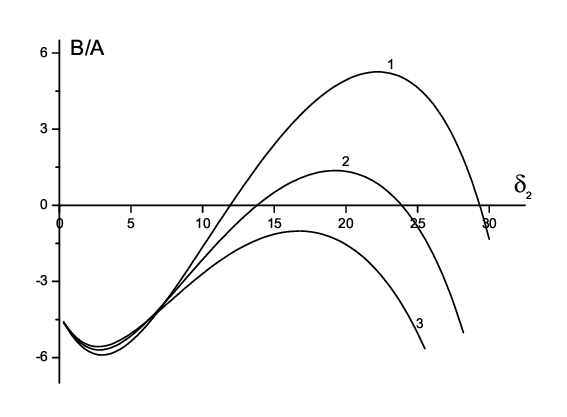}
\\Fig.4
\end{figure}

\begin{figure}\center
\includegraphics[height=6cm]{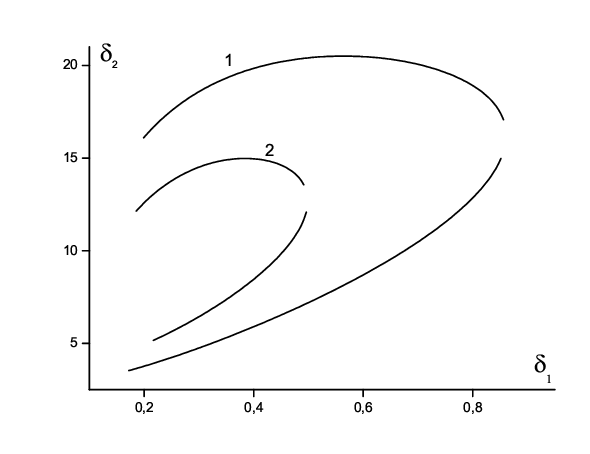}
\\Fig.5
\end{figure}

\begin{figure}\center
\includegraphics[height=6cm]{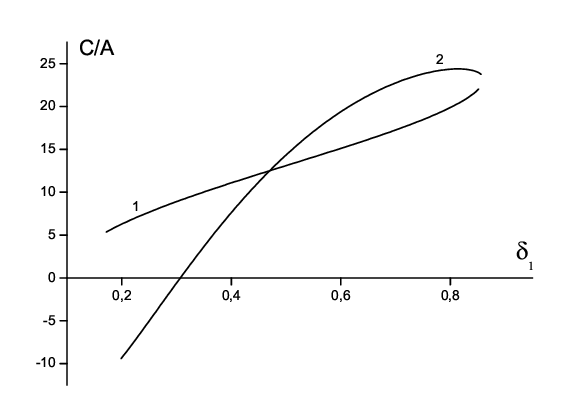}
\\Fig.6
\end{figure}

\end{document}